\documentclass[10pt,twocolumn,amssymb,nobibnotes,aps,prb]{revtex4-1}
\usepackage{graphicx}
\usepackage{amssymb}
\usepackage{amsmath}
\usepackage{xspace}
\usepackage{url}
\usepackage{color}
\begin{document}
\definecolor{dg}{rgb}{0.00, 0.70, 0.40}
\newcommand{\nag}{\phantom{\dag}}
\title{Thermoelectric effects in molecular quantum dots with contacts}
\author{T. Koch$^1$, J. Loos$^2$, H. Fehske$^1$}
\affiliation{$^1$Institut f{\"u}r Physik,
            Ernst-Moritz-Arndt-Universit{\"a}t Greifswald,
             DE-17489 Greifswald, Germany\\
             $^2$Institute of Physics, Academy of Sciences of the Czech Republic, CZ-16200 Prague, Czech Republic
}
\date{\today}
\begin{abstract}
We consider the steady-state thermoelectric transport through a vibrating molecular quantum dot that is contacted to macroscopic leads. For moderate electron-phonon interaction strength and comparable electronic and phononic timescales, we investigate the impact of the formation of a local polaron on the thermoelectric properties of the junction. We apply a variational Lang-Firsov transformation and solve the equations of motion in the Kadanoff-Baym formalism up to second order in the dot-lead coupling parameter. We calculate the thermoelectric current and voltage for finite temperature differences in the resonant and inelastic tunneling regimes. For a near resonant dot level, the formation of a local polaron can boost the thermoelectric effect because of the Franck-Condon blockade. The line shape of the thermoelectric voltage signal becomes asymmetrical due to the varying polaronic character of the dot state and in the nonlinear transport regime, vibrational signatures arise.
\end{abstract}
\maketitle
%
%
\section{Introduction}

Molecular junctions are electronic devices that consist of an organic quantum dot that is contacted by two macroscopic leads. Modern nanotechnology allows for the reliable fabrication of systems where the dots are single aromatic rings, molecular wires\cite{Malen2009b}, C$_{60}$ fullerenes\cite{Franke2012,Evangeli2013} or carbon nanotubes.\cite{Pop2005,*Leturcq2009} They are a promising candidates in the search for further miniaturization of electronic and thermoelectric devices.\cite{Galperin2007a,*Dubi2011}

Transport through such systems is determined by the discrete levels of the dot, whose position relative to the Fermi energy can be tuned, e.g., with the help of a third (gate) electrode.~\cite{Scheibner2008,*Song2009} The level broadening depends on the dot-lead coupling strength, which can be manipulated via the lead distance or through the choice of different metal-molecule anchoring groups.~\cite{Cheng2011} In addition to these two external parameters, electron-phonon (EP) interaction influences transport through molecular junctions: When it is occupied by charge carriers, the molecule may undergo structural changes or vibrations that correspond to the excitation of local optical phonons of considerable energy. They show up as vibrational signatures in the current-voltage characteristics of the device~\cite{Reed1997,*Park2000,Franke2012}.

A temperature difference between the leads induces a current of charge carriers across the junction. This thermoelectric effect is measured by recording, for constant temperature difference, the voltage bias necessary to cancel this current.~\cite{Reddy2007b} The quotient of the temperature difference and the thermovoltage, the so-called thermopower, can be used to probe the systems Fermi energy\cite{Baheti2008} and the vibronic structure of the molecule's state.\cite{Koch2004,Schaller2013} Most of the experimental findings are well understood within a linear response formulation of the thermopower.\cite{Ke2009} However, the applied temperature differences can be tens of degrees Kelvin, i.e. larger than the dot-lead coupling energy. Some features of the measured voltage histograms, such as side peaks and temperature dependent widths,\cite{Reddy2007b} are not accounted for in the linear theory. That is why recently, the discussion of the thermopower has been extended to the nonlinear regime.\cite{Leijnse2010,Bergfield2010,*Hsu2012} Then the question arises, how the thermoelectric coefficients can be generalized. Our approach in the present work is motivated by the experimental situation: For a given temperature bias we determine the thermovoltage numerically by minimizing the thermally induced charge current.

We base our calculations on the Anderson-Holstein model. Here, the organic molecule is represented by a single energy level and a local optical mode, which is linearly coupled to the electron on this level. The quantum dot is connected to two macroscopic leads, while the local mode is coupled to a phonon bath. The current between the leads is given by the interacting dot density of states.\cite{Meir1992} Based on such models, different methods have been applied, such as the numerical\cite{Jovchev2013,Eidelstein2013} and functional renormalization group\cite{Laasko2013}, rate equations,\cite{Beenakker1992,Koch2004} master equations\cite{Leijnse2010,Sanchez2011,*Juergens2013,Schaller2013} and nonequilibrium Keldysh Green functions\cite{Galperin2007b} to describe transport through the dot for small-to-large dot-lead coupling and weak-to-strong EP interaction.

In the antiadiabatic, strong EP coupling regime, typically a Lang-Firsov transformation\cite{Lang1962} is applied, based on the exact solution of the isolated dot. It predicts the formation of a local polaron, which reduces the effective dot-lead coupling exponentially. This could be beneficial for the thermoelectric response of the system.~\cite{Humphrey2002,Leijnse2010} For practical applications, however, a moderate level broadening is needed to ensure usable power output. Moreover, long electron residence times and strong EP interaction may lead to the accumulation of energy at the dot and, consequently, to its degradation.\cite{Schulze2008} Because of this, the regime of comparable electronic and phononic time scales and intermediate EP interaction becomes interesting. 

To account for the polaronic character of the dot-state away from the strong EP coupling, antiadiabatic limit, we use an approach that is based on a variational form of the Lang-Firsov transformation.\cite{LaMagna2007,*LaMagna2009} For the polaron problem, the variational Lang-Firsov approach has been proven to give reliable results in the whole electron-phonon coupling and phonon frequency regime, even in the most physically
difficult polaron crossover region.\cite{Fehske1994,*Fehske1997} Here, the polaron variational parameter is determined by minimizing the relevant thermodynamical potential.

In previous work,~\cite{Koch2011,*Koch2012} we considered the steady-state current response of the quantum dot to a finite voltage difference between the leads. In doing so, we assumed that the temperatures of the leads and of the phonon bath are all equal. The calculations in these papers were based on the Kadanoff-Baym formalism,~\cite{Kadanoff1962} which relies on the relation between the real-time response functions and the non-equilibrium Green functions of the complex time variables.

The present work will focus on the thermoelectric effects induced by a finite temperature difference between the leads. 
Section~\ref{SECmodel} introduces the model and the variational ansatz we employ to describe a vibrating quantum dot that is coupled to a phonon heat bath and two macroscopic leads at different temperatures. Because for such a setup the temperature is not constant throughout the system, the reasonings of our earlier approach~\cite{Koch2011,*Koch2012} have to be modified. That is why in Sec.~\ref{SECresponse} we generalize the Kadanoff-Baym method such  that the steady-state equation of the response functions, as well as their formal solution deduced in Ref.~\onlinecite{Koch2011}, are applicable to the present case. In Sec.~\ref{SECself} we then derive an approximation for the polaronic self-energy, and Sec.~\ref{SECcurrent} provides the relation between the polaronic and electronic spectral functions, the latter of which enters the current formula. In Sec.~\ref{SECvar} we derive the thermodynamic potential which will be used to determine the variational parameter numerically. Section~\ref{SECnumerics} presents our numerical results. The main conclusions  and future prospects can be found
in Sec.~\ref{SECsummary}.
%
%
\section{Theory}\label{SECtheory}
%
%
\subsection{Model and variational ansatz}\label{SECmodel}
\begin{figure}[t]
\begin{center}
	\includegraphics[width=0.9\linewidth]{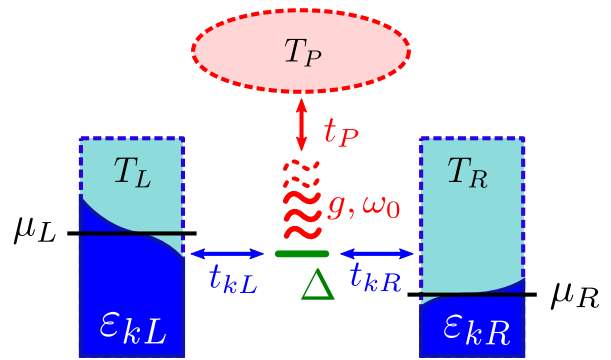}
\caption{(Color online) Sketch of the quantum dot model. The electronic level $\Delta$ is coupled to two macroscopic leads, each in its own thermal equilibrium with different temperatures $T_{L,R}$ and chemical potentials $\mu_{L,R}$. The dot electrons interact with an optical phonon mode of energy $\omega_0$ that is coupled to a phonon bath at temperature $T_P$.}
\label{FIG1}
\end{center}
\end{figure}

We consider a molecular quantum dot in the three-terminal configuration depicted in Fig.~\ref{FIG1}, which will be modelled by the following Hamiltonian:
\begin{align}
 H  &=  (\Delta-\mu) d^\dag d^{\nag} -\; g \omega_0 d^\dag d ( b^{\dag} + b) + \omega_0  b^{\dag} b  \label{EQUhamiltonianstart}\\
&+ \sum_{k,a} (\varepsilon_{ka}^{\nag}-\mu) c_{ka}^{\dag} c_{ka}^{\nag} - \frac{1}{\sqrt{N}}\sum_{k, a}\left(t_{ka}d^{\dag}c_{ka}^{\nag} + t_{ka}^\ast c_{ka}^{\dag}d\right) .\nonumber
\end{align}
Here, the quantum dot is represented by a single energy level $\Delta$ with the fermionic operators $d^{(\dag)}$. It interacts with a local optical phonon mode $b^{(\dag)}$ of energy $\omega_0$, where the EP coupling strength is given by the so-called polaron binding energy,
\begin{align}
\varepsilon_p=g^2\omega_0\;.
\end{align} 
By the last term in Eq.~(\ref{EQUhamiltonianstart}), the dot is coupled to left ($a=L$) and right ($a=R$) leads, each of which contains $N$ free electrons with the energies $\varepsilon_{ka}$ and the corresponding fermionic operators $c_{ka}^{(\dag)}$, respectively. In equilibrium, the dot-lead system is characterized by a common chemical potential $\mu$ and a global temperature $T$. 

In the nonequilibrium situation, a voltage difference between the leads is described by adding to (\ref{EQUhamiltonian}) the term
\begin{align}\label{EQUhint}
H_{\mathrm{int}}&=\sum_{a}U_a\sum_{k}c_{ka}^\dag c_{ka}^{\nag}
\end{align} 
with the voltage bias
\begin{align}\label{EQUdefPhi}
\Phi=(U_L-U_R)/e \;,
\end{align}
where $e<0$ is the electron charge. In addition, we consider a temperature difference between the macroscopic leads, whereby each lead is supposed to stay in its own thermal equilibrium with the temperatures $T_L$ and $T_R$, respectively. We also suppose that the local oscillator is coupled to its own heat bath, which has the temperature $T_{P}$. In accordance with Entin-Wohlman~\cite{Entin2010} we assume that the coupling to this heat bath (indicated as $t_P$ in Fig.~\ref{FIG1}) far exceeds the EP-coupling strength, so that the phonon population at the dot is given by the Bose-Einstein distribution 
\begin{align}
n_B(\omega) &= (e^{\beta_{P}\omega}-1)^{-1}\quad\textnormal{with}\quad \beta_{P}=(k_B T_{P})^{-1}\;.\label{EQUbosebath}
\end{align}
In the following calculations, we suppose that $T_P=T_R=T$ and $T_L\neq T$ in general. Moreover, all energies will be taken with respect to the equilibrium chemical potential, i.e., $\mu=0$. We raise the voltage bias symmetrically around $\mu$, i.e., $\mu_L=-e\Phi/2$ and $\mu_R=e\Phi/2$. 
%

%
To account for the polaron localization at the dot, we apply to the Hamiltonian (\ref{EQUhamiltonianstart}) an incomplete Lang-Firsov transformation $S_\gamma$,~\cite{LaMagna2007,Koch2011,Koch2012} introducing the variational parameter $\gamma\in[0,1]$:
\begin{align}
\widetilde H&=S_\gamma^\dag H S_\gamma\;,\quad
S_\gamma=\exp\{\gamma g (b^{\dag}-b)d^\dag d\}\;. \label{EQUlft}
\end{align}
The transformed Hamiltonian then reads
\begin{align}
\widetilde H &=  \widetilde\Delta\, d^\dag d^{\nag} -C_{d}^{\nag} d^\dag d  + \omega_0  b^{\dag} b  \nonumber\\
&+\sum_{k,a}\varepsilon_{ka}c_{ka}^{\dag} c_{ka}^{\nag}-\sum_{k,a} \left(  C_{ka}^{\nag} d^{\dag}c_{ka}^{\nag}+ C_{ka}^\dag c_{ka}^{\dag}d\right)  \;,\label{EQUhamiltonian}
\end{align}
where the renormalization of the dot level and the interaction coefficients depends on the parameter $\gamma$:
\begin{align}
\widetilde\Delta&=\Delta-\varepsilon_p\gamma(2-\gamma)\;,\label{EQUdefeta}\quad \widetilde{g}=\gamma g\;,\\
C_{ka}&=\frac{t_{ka}}{\sqrt{N}}\,\mathrm{e}^{-\widetilde g (b^{\dag}-b)}\;,\quad C_d=g\omega_0(1-\gamma)(b^\dag+b)\;.\label{EQUdefinteraction}
\end{align}
Now $d$ and $b$ are the operators of dressed electrons (in analogy to polarons) and the shifted local oscillator.
The original electron and oscillator operators read $\widetilde d= \mathrm{exp}\{\gamma g (b^{\dag}-b)\}\,d$ and $\widetilde b=b+\gamma g d^\dag d$, respectively.
%
%
\subsection{Response functions and steady-state equations}\label{SECresponse}
In our previous works,~\cite{Koch2011,Koch2012} we considered the electron current response of a molecular quantum dot that was initially in equilibrium with the leads as well as with the phononic bath at a common, fixed temperature. Our calculations were based on the Kadanoff-Baym theory.~\cite{Kadanoff1962} At first, we summarize the aspects of this non-equilibrium response theory that are essential for the present setting. 

The relevant response functions are represented by the real-time Green functions of the dot-operators $d^{(\dag)}$ and the lead operators $c_{ka}^{(\dag)}$, e.g.
\begin{align}
g_{dd}(t_1,t_2;U)&=-\mathrm{i}\langle \mathcal{T} d_U(t_1)d_U^\dagger(t_2)\rangle\;, \label{EQUdefresponse}\\
g_{dd}^<(t_1,t_2;U)&=\mathrm{i}\langle  d_U^\dagger(t_2)d_U(t_1)\rangle \;,\label{EQUdefresponseless}\\
g_{dd}^>(t_1,t_2;U)&=-\mathrm{i}\langle  d_U(t_1)d_U^\dagger(t_2)\rangle \;.\label{EQUdefresponsegtr}
\end{align}
The functions $g_{cd}$ of the ``mixed'' operators $c_{ka}^{(\dag)}$ and $d^{(\dag)}$ are defined in an analogous way. In Eqs.(\ref{EQUdefresponse})--(\ref{EQUdefresponsegtr}) the time dependence of $d_U^{(\dagger)}$ is determined by $\widetilde H+H_{\mathrm{int}}$. The symbol $\mathcal{T}$ means the standard time-ordering operator so that the function $g_{dd}$ is equal to the functions $g_{dd}^<(t_1,t_2;U)$ and $g_{dd}^>(t_1,t_2;U)$ for $t_1<t_2$ and $t_1>t_2$, respectively.
The statistical average $\langle \cdots\rangle$ corresponds to the equilibrium state at the temperature $T$ before the disturbance was turned on.
Going to the interaction representation, the Heisenberg operators $d_U$ are expressed as
\begin{align}
d_U(t)&= V^{-1}(t)d(t)V(t)\;,\\
 V(t)&=\mathcal{T}_t\exp\left\{-\mathrm{i}\int_{-\infty}^{t}\mathrm{d}t^\prime \;H_{\mathrm{int}}(t^\prime)\right\} \;.
\end{align}
The real-time response functions may be deduced using the equations of motion for the nonequilibrium Green functions of the complex time variables $t=t_0-\mathrm{i}\tau$, $\tau\in[0,\beta]$, defined as
\begin{align}
G_{dd}(t_1,t_2;U,t_0)&=-\frac{\mathrm{i}}{\langle S \rangle}\langle \mathcal{T}_\tau d(t_1)d^\dagger(t_2)S\rangle \label{EQUdefG}\;,\\
S=\mathcal{T}_\tau & \exp\left\{-\mathrm{i}\int_{t_0}^{t_0-\mathrm{i}\beta}\mathrm{d}t \;H_{\mathrm{int}}(t)\right\} \;.\label{EQUexponential}
\end{align}
The time dependence of the operators in Eq.~(\ref{EQUdefG}) is determined by $\widetilde H$ only, while the external disturbance is explicit in the time-ordered exponential operator $S$. 
The operation $\mathcal{T}_\tau$ orders the operators by the imaginary parts of the times $t_1$ and $t_2$, so that $G_{dd}(t_1,t_2;U,t_0)=G_{dd}^>(t_1,t_2;U,t_0)$ for $\mathrm{i}(t_1-t_2)>0$ and $G_{dd}(t_1,t_2;U,t_0)=G_{dd}^<(t_1,t_2;U,t_0)$ for $\mathrm{i}(t_1-t_2)<0$. To find the relation between the functions $G_{dd}^{\lessgtr}$ and $g_{dd}^{\lessgtr}$, the function $G_{dd}$ is considered for $\mathrm{i}(t_1-t_2)<0$:
%
\begin{align}
G_{dd}^<(t_1,t_2;U,t_0)= &\frac{\mathrm{i}}{\langle U(t_0,t_0-\mathrm{i}\beta)\rangle}\times \\
\Big \langle & U(t_0,t_0-\mathrm{i}\beta) U^{-1}(t_0,t_2)d^{\dag}(t_2) U(t_0,t_2)\nonumber\\
& U^{-1}(t_0,t_1) d(t_1) U(t_0,t_1) \Big \rangle \nonumber
\end{align}
%
where
\begin{align}
U(t_0,t)&=\mathcal{T}_\tau\exp\left\{-\mathrm{i}\int_{t_0}^{t}\mathrm{d}t^\prime \;H_{\mathrm{int}}(t^\prime)\right\} \;.
\end{align}
The continuation of $U(t_0,t)$ and $V(t)$ to analytic functions of the time variables leads to the identification of $U(t_0,t)$ with $V(t)$ in the limit $t_0\to-\infty$. Consequently, the connection of the analytic functions $G_{dd}^\lessgtr$ and $g_{dd}^\lessgtr$ is given by
\begin{align}
\lim_{t_0\to-\infty} G_{dd}^{\lessgtr}(t_1,t_2;U,t_0) &= g_{dd}^{\lessgtr}(t_1,t_2;U)\label{EQUlimitlessgtr}\;,
\end{align}
with similar relations for the functions $G_{cd}^{\lessgtr}$ and $g_{cd}^{\lessgtr}$.

It is evident that the derivation of Eq.~(\ref{EQUlimitlessgtr}) outlined above  does not refer to some special properties of the statistical ensemble, nor to the physical meaning of $\beta$. In this way, it is possible to generalize the definition (\ref{EQUdefG}) for $G_{dd}$ assuming the mean value $\langle \cdots \rangle$ to be unspecified and the complex-time variable to be defined in the interval $t\in [t_0,t_0-\mathrm{i}\sigma]$, where the time-ordering parameter $\sigma$ has no specific physical meaning. We assume that before the disturbance (\ref{EQUhint}) was turned on, the system was in a steady state with the temperatures $T_L$, $T_R$ of the left and right leads, and $T_P$ of the phonon bath. The function $G_{dd}$ defined in this way does not have the properties of the temperature ($\beta$) dependent Green function in Eq.~(\ref{EQUdefG}), it rather represents a functional of the ordered operators which we use to determine the real-time response funcitons $g_{dd}$. 

To do this, we define, in analogy to the self energy, the function $\Sigma_{dd}(t_1,t_2;U,t_0)$ by the equation
\begin{align}
\left[G_{dd}^{(0)-1}(t_1,\bar t)-\Sigma_{dd}(t_1,\bar t;U,t_0)\right]\bullet& G_{dd}  (\bar t,t_2;U,t_0)\nonumber \\
&=\delta(t_1-t_2)\;,\label{EQUdyson}
\end{align}
with the inverse zeroth-order function 
\begin{align}
G_{dd}^{(0)-1}(t_1,t_2)=\left ( \mathrm{i}\frac{\partial}{\partial t_1} - \widetilde \Delta \right ) \delta (t_1-t_2)\;.\label{EQUinverseG0}
\end{align}
In Eq.~(\ref{EQUdyson}) the matrix multiplication ``$\bullet$'' is defined by the integration $\int_{t_0}^{t_0-\mathrm{i}\sigma}\mathrm{d}\bar t\cdots$ containing the time-ordering parameter $\sigma$. The $\delta$-function of complex arguments is understood with respect to this integration. 
The inverse function to $G_{dd}$ is given as $G_{dd}^{-1}(t_1,t_2;U,t_0) =  G_{dd}^{(0)-1}  (t_1,t_2)-\Sigma_{dd}(t_1,t_2;U,t_0)$. The deduction of the steady-state equations for the real-time functions $g^<(t_1,t_2;U)$ and $g^>(t_1,t_2;U)$ by the limiting procedure $t_0\to-\infty$ is analogous to that given in Ref.~\onlinecite{Koch2011}. 
Defining the Fourier transformations according to Kadanoff-Baym,\cite{Kadanoff1962}
\begin{align}
g_{dd}^{\lessgtr} (\omega;U) &= \mp\mathrm{i}\int_{-\infty}^{\infty} \mathrm{d}t\;g_{dd}^{\lessgtr}(t;U)e^{\mathrm{i}\omega t}
\end{align}
and similarly for $\Sigma^{\lessgtr}(\omega;U$), the solution of the steady-state equations may be written as follows:
\begin{align}
A(\omega;U) &= g_{dd}^>(\omega;U)+g_{dd}^<(\omega;U)\;,\label{EQUdefa} \\
g_{dd}^<(\omega;U)&=A(\omega;U)\bar f(\omega;U)\;,\label{EQUgless}\\
\bar f (\omega;U)&= \frac{\Sigma_{dd}^<(\omega;U)}{\Gamma(\omega;U)}\;,\label{EQUbarf}\\
\Gamma(\omega;U)&=\Sigma_{dd}^>(\omega;U)+\Sigma_{dd}^<(\omega;U) \;,\label{EQUdefGamma}
\end{align}
with the non-equilibrium polaronic spectral function
\begin{align}
A(\omega;U)&=\dfrac{\Gamma(\omega;U)}{\left(\omega-\widetilde\Delta-\mathcal{P}\int\frac{\mathrm{d}\omega^\prime}{2\pi}\;\frac{\Gamma(\omega^\prime;U)}{\omega-\omega^\prime}\right)^2+\left(\frac{\Gamma(\omega;U)}{2}\right)^2}\;.\label{EQUafrac}
\end{align}
%
%
\subsection{Self energy}
\label{SECself}
According to the preceeding section, the concrete form of the steady-state solution for the special choice of interactions is determined by the functions $\Sigma_{dd}^{\lessgtr}$. To find an explicit expression for $\Sigma_{dd}^{\lessgtr}$, we start with the equations of motion for $G_{dd}$ and $G_{cd}$, which are given by the commutators of the operators $d^{(\dag)}$ and $c_{ka}^{(\dag)}$ with $\widetilde H$. 
As a purely formal device, we add to $H_\mathrm{int}$ in Eq.~(\ref{EQUhint}) the interaction with fictitious external fields $\{V\}$. The equations of motion of $G_{dd}$ and $G_{cd}$ are then expressed by means of the functional derivatives of $\Sigma_{dd}$ with respect to these fields.  The resulting equations for $\Sigma_{dd}^{\lessgtr}$ are solved iteratively.
We then let $\{V\}\to 0$ and perform the limit $t_0\to -\infty$. 
In the following calculations we will use the self-energy function after the first iteration step:\cite{Koch2011}
\begin{align}
\Sigma_{dd}^{(1)\lessgtr}  (t_1,t_2;U)&=\sum_{k,a}|\langle C_{ka}\rangle |^2 \; g_{cc}^{(0)\lessgtr}(k,a;t_1,t_2;U) \; \nonumber\\
&\quad\times \Big\{I_0(\kappa) + \sum_{s\ge 1} I_s(\kappa) 2\sinh(s\theta)\nonumber\\
&\quad\times\Big [ (n_B(s\omega_0)+1)\mathrm{e}^{\pm\mathrm{i}s\omega_0(t_1-t_2)} \nonumber\\
&\quad+ n_B(s\omega_0)\mathrm{e}^{\mp\mathrm{i}s\omega_0(t_1-t_2)} \Big ] \Big \}\;, \label{EQUSigma1lessgtr}
\end{align}
where we have defined
\begin{align}
\theta&=\frac{1}{2}\omega_0\beta_P\;,\quad\kappa=\frac{\widetilde g^2}{\sinh\theta}\;,\\ 
&I_s(\kappa)=\sum_{m=0}^{\infty}\frac{1}{m!(s+m)!}\left(\frac{\kappa}{2}\right)^{s+2m}\;.\label{EQUdefthetakappa}
\end{align}
The Bose-function $n_B$ in (\ref{EQUSigma1lessgtr}) contains the phonon-bath temperature $T_P$ according to Eq.~(\ref{EQUbosebath}), while the zeroth-order functions of the leads depend on the different temperatures $T_R$ and $T_L$:
\begin{align}
g_{cc}^{(0)<}(k,a;t_1,t_2;U)&=\mathrm{i}e^{-\mathrm{i}\varepsilon_{ka}t}f_a(\varepsilon_{ka}+U_a)\;,\\
g_{cc}^{(0)>}(k,a;t_1,t_2;U)&=-\mathrm{i}e^{-\mathrm{i}\varepsilon_{ka}t}[1-f_a(\varepsilon_{ka}+U_a)]\;,
\end{align}
with the lead Fermi-functions
\begin{align}
f_a(\omega)&=\Big(e^{\beta_{a}\omega}+1\Big)^{-1}\;,\quad \beta_{a}=(k_B T_a)^{-1}\label{EQUfermileads}\;.
\end{align}
The Fourier-transformation of Eq.~(\ref{EQUSigma1lessgtr}) leads to
\begin{align}
\Sigma_{dd}^{(1)<}&(\omega;U)= \sum_a \Big \{ I_0(\kappa) \widetilde\Gamma^{(0)}_a(\omega)f_a(\omega+U_a)\nonumber\\
&+ \sum_{s\ge 1} I_{s}(\kappa) 2 \sinh (s \theta)\label{EQUSigmalessfourier1}\\
&\times \Big [ \widetilde\Gamma^{(0)}_a(\omega-s\omega_0) n_B(s\omega_0)f_a(\omega-s\omega_0+U_a) \nonumber\\
&+  \widetilde\Gamma^{(0)}_a(\omega+s\omega_0)(n_B(s\omega_0)+1) f_a(\omega+s\omega_0+U_a)  \Big ] \Big \} \;, \nonumber
\end{align}
with the renormalized dot-lead coupling function
\begin{align}
\widetilde\Gamma_a^{(0)}(\omega) & = \mathrm{e}^{-\widetilde g^2\coth\theta}\Gamma_a^{(0)}(\omega),\label{EQUtildegamma} 
\end{align}
which depends on the lead density of states:
\begin{align}
\Gamma^{(0)}_a(\omega)&=2\pi |t_a(\omega)|^2\frac{1}{N}\sum_k\delta(\omega-\varepsilon_{ka})\;.
\end{align}
From Eq.~(\ref{EQUSigmalessfourier1}), $\Sigma_{dd}^{(1)>}(\omega;U)$ results by interchanging $n_B\leftrightarrow (n_B+1)$ and $f_a\leftrightarrow (1-f_a)$. Note that Eq.~(\ref{EQUSigmalessfourier1}) is similar to our result for the self-energy in Ref.~\onlinecite{Koch2011}, but now $f$ is replaced by the individual lead Fermi-functions $f_a$ that were defined in (\ref{EQUfermileads}). 
If we insert the approximation (\ref{EQUSigmalessfourier1}) into Eq.~(\ref{EQUdefGamma}), we find
\begin{align}
\Gamma^{(1)}&(\omega;U) = \Gamma_{L}^{(1)}(\omega;U) + \Gamma_{R}^{(1)}(\omega;U)\;,\label{EQUgamma1}\\[0.2cm]
\Gamma_{a}^{(1)}&(\omega;U)= I_{0}(\kappa)  \widetilde\Gamma^{(0)}_a(\omega)\label{EQUgamma1a}\\
& + \sum_{s\ge 1} I_{s}(\kappa)2\sinh(s\theta)\nonumber\\
& \times \Big[  \widetilde\Gamma^{(0)}_a(\omega-s\omega_0)\Big(n_B(s\omega_0)+1-f_a(\omega+U_a-s\omega_0)\Big) \nonumber\\
& + \widetilde\Gamma^{(0)}_a(\omega+s\omega_0)\Big (n_B(s\omega_0)+f_a(\omega+U_a+s\omega_0) \Big)\Big] \;.\nonumber
\end{align}
From the functions $\Gamma^{(1)}(\omega;U)$ and $\Sigma_{dd}^{(1)<}(\omega;U)$, the first order spectral function $A^{(1)}(\omega;U)$, the distribution function $\bar f^{(1)}(\omega;U)$ and the response functions $g_{dd}^{(1)\lessgtr}(\omega;U)$ follow according to Eqs.~(\ref{EQUafrac}), (\ref{EQUbarf}) and (\ref{EQUgless}), respectively. 
%
%
%
\subsection{Electron current}\label{SECcurrent}
The operator of the particle current from lead $a$ to the dot reads
\begin{align}
\hat J_a
&=\frac{\mathrm{i}}{\sqrt{N}} \sum_k \left [ t_{ka} \widetilde d^\dag c_{ka}^{\nag} - t_{ka}^\ast c_{ka}^\dag \widetilde d \right ]\;.\label{EQUcurrentop}
\end{align}
Its mean value $J_a=\langle \hat J_a \rangle$ is given by the real-time response functions $\widetilde g_{cd}$ of the electron operators, which are defined in analogy to Eqs.~(\ref{EQUdefresponseless}) and (\ref{EQUdefresponsegtr}), e.g.
\begin{align}
\mathrm{i}\langle \widetilde d^\dag c_{ka}^{\nag} \rangle &= \widetilde g_{cd}^<(k,a;t_1,t_1;U) \label{EQUmeandc}\,.
\end{align}
We determine $\widetilde g_{cd}$ based on the equation of motion of the corresponding function $\widetilde G_{cd}(k,a;t_1,t_2;U,t_0)$ of complex-time variables:
\begin{align}
\widetilde G_{cd} & (k,a;t_1,t_2;U,t_0) = \nonumber\\
&-\frac{ t_{ka}^\ast}{\sqrt{N}} G_{cc}^{(0)}(k,a;t_1,\bar t;U) \bullet\widetilde G_{dd} (\bar t,t_2;U,t_0)\;.\label{EQUcomplexeom1}
\end{align}
The current $J_a$ results as
\begin{align}
J_a &= \int_{-\infty}^{\infty}\frac{\mathrm{d}\omega}{2\pi}\; \Gamma_a^{(0)}(\omega)\nonumber\\
&\times\left \{ f_a(\omega+U_a)  \widetilde A^{(1)}(\omega;U) -\widetilde g_{dd}^{(1)<}(\omega;U)  \right \}\;,\label{EQUcurrenta}
\end{align}
where $\widetilde A^{(1)}$ and $\widetilde g_{dd}^{(1)<}$ are the electronic dot spectral function and response function. In the steady state regime our approximation conserves the particle current: $J_L+J_R=0$. In the following, identical leads are assumed and we work in the wide band limit where $\Gamma_L^{(0)}(\omega)=\Gamma_R^{(0)}(\omega)\equiv\Gamma_0$. Then the total particle current through the dot, $J=(J_L-J_R)/2$, is given by 
\begin{align}
J&=\frac{\Gamma_0}{2}\int_{-\infty}^{\infty}\frac{\mathrm{d}\omega}{2\pi}\,\,\widetilde A^{(1)}(\omega;U) \left[ f_L(\omega+U_L) -  f_R(\omega+U_R) \right ]\;.\label{EQUcurrent} 
\end{align}
Based on Eq.~(\ref{EQUcurrent}), we determine the electron current through the quantum dot numerically. We note that in the present work we are only interested in the current response to finite voltage biases and temperature differences between the leads. The generation of an electronic response merely by the interaction with a hot phonon bath would require different, energy-dependent densities of states in the leads, as was shown in Ref.~\onlinecite{Entin2010}. In our calculations however, we consider identical lead densities of states. That is why $J=0$ for all $T_P$, as long as $T_L=T_R$ and $\Phi=0$.

To find a relation between the polaronic and electronic functions, we decouple the fermionic and bosonic degrees of freedom in the electronic response function as is customary,\cite{Lundin2002,*Zhu2003,Galperin2006}
\begin{align}
\widetilde g_{dd}^{<}  (t_1,t_2;U)&\approx g_{dd}^{<}(t_1,t_2;U)\,\langle e^{-\widetilde g(b^\dag-b)t_2}e^{\widetilde g(b^\dag-b)t_1}\rangle\;, \label{EQUDefGelec}
\end{align}
As stated in Sec.~\ref{SECmodel} the local oscillator is supposed to be strongly coupled to the heat bath. Accordingly we neglect the influence of the EP interaction on the dynamics of the phonon subsystem and evaluate the bosonic correlation function in (\ref{EQUDefGelec}) using Eq.~(\ref{EQUbosebath}). The first order electronic response functions then read
\begin{align}\label{EQUelectronicg}
\widetilde g_{dd}^{(1)\lessgtr} (\omega;U)&=\mathrm{e}^{-\widetilde g^2\coth\theta}\Big\{ I_{0}(\kappa)g_{dd}^{(1)\lessgtr}(\omega;U)\nonumber\\
&+ \sum_{s\ge 1} I_{s}(\kappa)2\sinh(s\theta)\\
&\times\Big( [1+n_B(s\omega_0)]g_{dd}^{(1)\lessgtr}(\omega\pm s\omega_0;U) \nonumber\\
&+ n_B(s\omega_0)g_{dd}^{(1)\lessgtr}(\omega\mp s\omega_0;U)  \Big ) \Big\}\;.\nonumber
\end{align}
The corresponding electronic spectral function $\widetilde A^{(1)}(\omega;U)$ follows according to the steady-state equation (\ref{EQUdefa}) as
\begin{align}\label{EQUelectronic}
\widetilde A&^{(1)}  (\omega;U)=\widetilde g_{dd}^{(1)<}(\omega;U)+\widetilde g_{dd}^{(1)>}(\omega;U)=\\[0.1cm]
&\quad\mathrm{e}^{-\widetilde g^2\coth\theta}\Big\{ I_{0}(\kappa)A^{(1)}(\omega;U) + \sum_{s\ge 1} I_{s}(\kappa)2\sinh(s\theta) \nonumber \\
&\quad\times\Big( [n_B(s\omega_0)+\bar f^{(1)}(\omega+s\omega_0;U)]A^{(1)}(\omega+ s\omega_0;U) \nonumber \\
&\quad+ [n_B(s\omega_0)+1-\bar f^{(1)}(\omega-s\omega_0;U)]A^{(1)}(\omega- s\omega_0;U)\Big ) \Big\}\;.\nonumber
\end{align}
%
%
%
\subsection{Variational procedure}\label{SECvar}
To determine the optimal variational parameter $\gamma_{min}$, we have to minimize the relevant thermodynamic potential. This poses a problem, since for finite $\Delta T$, the effective temperature determining the statistics of the dot electron is not known. It will be given by the constitution of the steady state due to the coupling of the dot with the surroudings. 
In the present paper, we suppose $\gamma$ to be mainly determined by the EP interaction terms contained in Eq.~(\ref{EQUhamiltonianstart}), since the variational parameter was introduced to characterize the polaron-like quasilocalization of the dot electron. Therefore we assume the equilibrium thermodynamic potential of the system, before the temperature- and voltage differences were turned on, to be a reasonable first approximation for the variational function. The use of the well-known ``$\lambda$ trick'' to take the interaction terms from $\widetilde H$  into account results in\cite{Kadanoff1962} 
\begin{align}\label{EQUthermpot}
\Omega&=-\frac{1}{\beta}\ln (1+\mathrm{e}^{-\widetilde\Delta\beta}) \nonumber\\
&+ 2 \int_{0}^{1}\frac{1}{\lambda} \int \frac{\mathrm{d} \omega}{2\pi} (\omega-\widetilde\Delta) A_\lambda^{(1)}(\omega) f(\omega)\;.
\end{align}
The function $A_\lambda^{(1)}$ represents the first order equilibrium spectral function if the interaction coefficients in Eq.~(\ref{EQUhamiltonian}) are multiplied by the factor $\lambda$. If we write $A_\lambda^{(1)}$ in terms of $\Gamma_\lambda^{(1)}=\lambda^2 \Gamma^{(1)}$, we can carry out the $\lambda$-integration and find
\begin{align}
\Omega& = - \int_{-\infty}^{\infty} \frac{\mathrm{d} \omega}{\pi} f(\omega) \Big \{   \frac{\pi}{2} \nonumber\\
&+  \arctan \Big ( \frac{\omega-\widetilde\Delta-\mathcal{P}\int \frac{\mathrm{d}\omega^\prime}{2\pi}\;\frac{\Gamma^{(1)}(\omega^\prime)}{\omega-\omega^\prime}} {\Gamma^{(1)}(\omega)/2} \Big)\Big \}\;.
\end{align}
Via $\widetilde\Delta$ and $\Gamma^{(1)}$, the thermodynamic potential $\Omega$ is a function of $\gamma$. For given parameters $\varepsilon_p$, $\Gamma_0$, $\omega_0$, $\Delta$ and $T$, the optimal variational parameter $\gamma_{min}$ will be determined in equilibrium, i.e., for $\Delta T=0$ and $\Phi=0$, by minimizing $\Omega$. We then keep $\gamma_{min}$ fixed and calculate the self energy, the spectral function and the resulting particle current for finite voltages or temperature differences between the leads.
%
%
\section{Numerical results}\label{SECnumerics}
Depending on the type of molecular junction, the energies of the characteristic vibrational modes range from the order of $10$meV in small molecules\cite{Zhitenev2002} to several $100$meV in $C_{60}$ molecules\cite{Franke2012}.
In the following calculations, the corresponding model parameter $\omega_0$ will be used as the unit of energy, i.e., we keep $\omega_0=1$ fixed and set $\hbar=1$, $|e|=1$ and $k_B=1$. We assume identical leads and work in the wide band approximation, so that the dot-lead coupling is determined by a single parameter: $\Gamma_L^{(0)}(\omega)=\Gamma_R^{(0)}(\omega)\equiv\Gamma_0$. Furthermore, we consider low equilibrium temperatures, $T/\omega_0\ll1$. Then, according to Eq.~(\ref{EQUtildegamma}), the renormalized dot-lead coupling parameter is approximately given as
\begin{align}
\widetilde\Gamma_0&\approx  \Gamma_0\,e^{-\gamma^2\varepsilon_p/\omega_0}\;.\label{EQUtildegamma2}
\end{align}
Usually, the ratio of the bare dot-lead coupling parameter $\Gamma_0$ to the phonon energy $\omega_0$ is used to determine whether the system is in the adiabatic ($\Gamma_0\gg\omega_0$) or anitadiabatic ($\Gamma_0\ll\omega_0$) regime.
Recently, Eidelstein {\it et al.}\cite{Eidelstein2013} argued that for strong EP interaction, i.e., $\varepsilon_p/\omega_0>1$, the antiadiabatic regime can be extended to $\Gamma_0\lesssim \omega_0$ as long the exponential renormalization of the dot-lead coupling parameter in (\ref{EQUtildegamma2}) is so strong that $\widetilde\Gamma_0\ll\omega_0$. In this case, the ``mobility'' of passing charge carriers is reduced so far that the local oscillator is still fast enough to adjust to them individually. Because then the physics is essentially those of the antiadiabatic regime, it can be called the ``extended antiadiabatic regime''.\cite{Eidelstein2013} Only when $\widetilde\Gamma_0$ approaches $\omega_0$ the system crosses over to the adiabatic regime. 

To investigate the latter situation, in the present work we consider comparable electronic and phononic timescales $\Gamma_0 \lesssim\omega_0$ and moderate EP coupling $\varepsilon_p\gtrsim \omega_0$. Since in our approach, the polaronic renormalization in Eq.~(\ref{EQUtildegamma2}) also depends on the value of the variational parameter $\gamma\in[0,1]$, we can interpolate between the extended adiabatic regime and the aforementioned crossover regime.

%
\subsection{Dot state in the crossover regime}\label{SECstate}
\begin{center}
\begin{figure}[t]
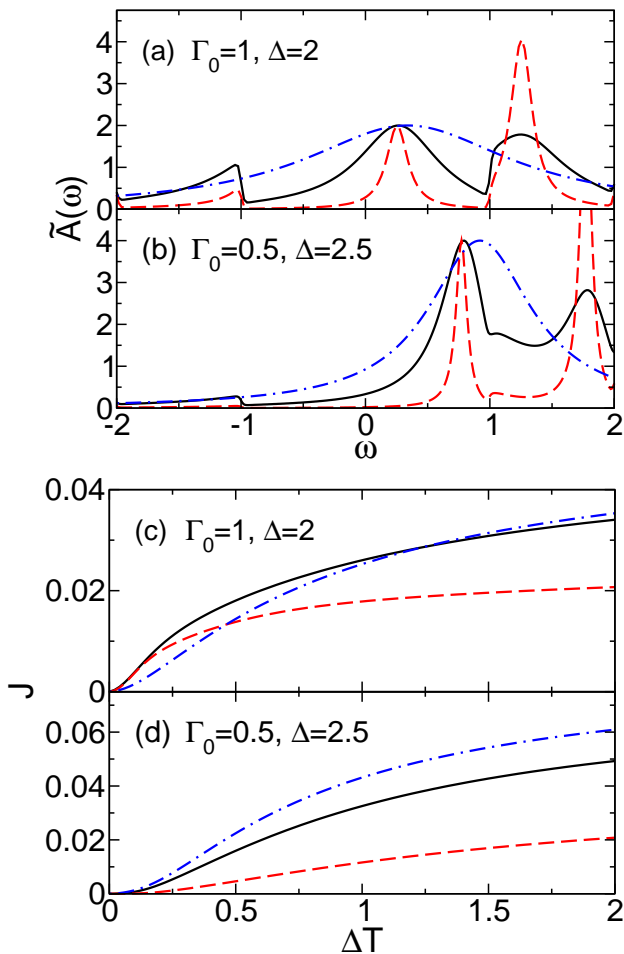

	\includegraphics[width=0.89\linewidth]{Fig2_a}\\[0.2cm]
	\hspace{-0.5cm}\includegraphics[width=0.95\linewidth]{Fig2_b}
\caption{(Color online) For $\omega_0=1$, $T=0.01$, and $\varepsilon_p=2$. (a) Electronic equilibrium spectral function for $\Gamma_0=1$, $\Delta=2$, $\gamma_{min}=0.59$, $\widetilde\Delta=0.33$ and $\widetilde\Gamma_0=0.5$. The variational calculation (black lines) is compared to the small polaron limit $\gamma=1$ (red dashed lines) and to the result for $\varepsilon_p=0$ (blue dot-dashed lines). In the latter two cases we set $\Delta$ in such a way that $\widetilde\Delta$ is the same as in the variational calculation. (b) Same as (a), but for $\Gamma_0=0.5$, $\Delta=2.5$, $\gamma_{min}=0.54$, $\widetilde\Delta=0.91$ and $\widetilde\Gamma_0=0.28$.  (c) and (d) For the same parameters as in (a) and (b), thermally induced current $J$ as a function of the temperature difference $\Delta T$ between the leads.
}
\label{FIG2}
\end{figure}
\end{center}
To understand the thermoelectric response of the dot we first investigate the dots electronic spectral function in equilibrium. Thereby we keep the parameters $\omega_0=1$, $T=0.01$, $\Phi=0$, $\Delta T=0$, and $\varepsilon_p=2$ fixed and consider two parameter sets for $\Gamma_0$ and $\Delta$.

The first set is $\Gamma_0=1$ and $\Delta=2$, which means that the electronic and phononic subsystems react on a similar timescale and the dot level acts as a tunneling barrier between the leads. For these parameters our variational calculation yields an optimal $\gamma_{min}=0.59$. As a consequence, the effective dot level given in Eq.~(\ref{EQUdefeta}) is lowered to near resonance, $\widetilde\Delta=0.33$. Moreover, the effective dot-lead coupling is reduced by about half, $\widetilde\Gamma_0\approx 0.5$, and we are in the aforementioned crossover regime.\cite{Eidelstein2013}

For the second parameter set we, reduced the dot-lead coupling to $\Gamma_0=0.5$ (meaning the phononic subsystem is the faster one) and raised the bare dot level to $\Delta=2.5$. This results in $\gamma_{min}=0.54$, $\widetilde\Gamma_0=0.28$, and $\widetilde\Delta=0.91$, so that the renormalized dot level is still far from resonance.  

The electronic spectral functions for the first and second parameter set are presented as the black curves in Figs.~\ref{FIG2}(a) and \ref{FIG2}(b), respectively. We compare them to two limiting cases: the red dashed curves represent the small polaron limit, where we set $\gamma=1$ instead of the optimal $\gamma_{min}$. Then $S_\gamma$ in Eq.~(\ref{EQUlft}) corresponds to the complete Lang-Firsov transformation. In this case, we have $\widetilde\Gamma_0=0.14$ and $\widetilde\Gamma_0=0.07$, respectively, and the system is described as being in the extended antiadiabatic regime. The blue dot-dashed curves follow from setting $\varepsilon_p=0$, which represents a rigid quantum dot without EP interaction. In both limiting cases we set $\Delta$ in such a way, that $\widetilde\Delta$ is the same as in the respective variational calculation.

In general, for $\varepsilon_p=0$, the electronic spectral function features a single broad band centered at $\omega=\Delta$. From setting $g=0$ in Eqs.~(\ref{EQUgamma1}) and (\ref{EQUafrac}), we see that it has a Lorentzian shape and its width is given by the bare dot-lead coupling parameter $\Gamma_0$. 
For finite EP interaction and $\gamma=1$, the polaronic character of the dot state is signalled by the appearance of several narrow side bands, which are given by the terms with $s\ge1$ in Eq.~(\ref{EQUelectronic}). These side bands represent the emission and absorption of optical phonons by the incident electrons. Their width is determined mainly by the small parameter $\widetilde\Gamma_0$ and their maxima are located multiple integers of $\omega_0$ away from the central (zero-phonon) peak. 

Note however, that the phonon peak corresponding to $\omega=\widetilde\Delta-\omega_0$ is suppressed in Figs.~\ref{FIG2}(a) and \ref{FIG2}(b), and only a small shoulder at $\omega=-\omega_0$ remains. This is a consequence of Pauli blocking and can be understood from Eqs.~(\ref{EQUSigmalessfourier1}) and (\ref{EQUgamma1a}); for small $\Delta T$, we have $n_B(s\omega_0)\approx0$ and the Fermi functions of both leads are nearly step-like. Then the third and fourth lines in Eqs.~(\ref{EQUSigmalessfourier1}) and (\ref{EQUgamma1a}) do not contribute in the region $\omega\in[-\omega_0+|\Phi|/2,+\omega_0-|\Phi|/2]$. From Eq.~(\ref{EQUbarf}) we see that $\bar f=(f_L+f_R)/2$ in this region and, consequently, the third and fourth line in Eq.~(\ref{EQUelectronic}) do not contribute to the spectral function. This ``floating'' of the phonon bands has been discussed for linear electric transport through molecular quantum dots.\cite{Mitra2004} As we will see, it is also the reason why the low temperature thermoelectric response is determined by the shape and position of the zero-phonon peak alone.

The variational dot state, due to the moderate renormalization, features broad, overlapping side bands. Still, we find a considerable shift of spectral weight to higher energies and a suppression of the zero-phonon peak. The resulting reduction of the low-energy tunneling rate is a main consequence of the polaron formation and is known as the Franck-Condon blockade.\cite{Jovchev2013}

\subsection{Thermally-induced current}\label{SECjtherm}

In the following we calculate the particle current through the junction that results from a finite temperature difference $\Delta T$ between the leads.
Let us again consider the two sets of parameters used in Sec.~\ref{SECstate}. Keeping the respective variational parameters $\gamma_{min}$ fixed we increase the temperature difference $\Delta T$, whereby $T_L=T+\Delta T$ and $T_P=T_R=T$. This induces a net particle current through the junction, which is depicted as a function of $\Delta T$ in Figs.~\ref{FIG2}(c) and \ref{FIG2}(d). For the first parameter set, we find that the EP interaction enhances the thermally induced current for small temperature differences. However, it reduces the current throughout the temperature range for the second parameter set. 

This can be understood from Eq.~(\ref{EQUcurrent}): the net current through the dot depends on the electronic nonequilibrium spectral function and the difference of the Fermi distribution functions of the left and right lead. For zero voltage bias and small temperature differences, the term $f_L(\omega)-f_R(\omega)$ in Eq.~(\ref{EQUcurrent}) differs from zero only in the narrow region $\omega\in[-\Delta T,+\Delta T]$ and changes sign at $\omega=0$. Physically, this means that the temperature difference between the leads induces a flow of hot charge carriers from the left to the right lead, since $f_L(\omega)-f_R(\omega)$ is positive for $\omega>0$. This current, however, is compensated by a counterflow of cold carriers at $\omega<0$, where $f_L(\omega)-f_R(\omega)$ is negative.  Since $\widetilde\Delta>0$ in our calculations, dot states above the Fermi energy have more spectral weight than the states below the Fermi energy. The net particle current is positive, i.e., it goes from the left to the right lead. The magnitude of the current is determined by the relative weight of the two flows, i.e., by the slope of the dot spectral function around the Fermi energy.

That is why the effect of the polaron formation on the thermally induced current crucially depends on the specific values of $\widetilde\Delta$ and $\widetilde\Gamma_0$. For the first parameter set, the dot level $\widetilde\Delta=0.33$ lies near the Fermi surface. In this situation, the polaronic renormalization of the effective dot-lead coupling increases the slope of $\widetilde A(\omega)$ at $\omega=0$, as can be seen in Fig.~\ref{FIG2}(a). Consequently, in Fig.~\ref{FIG2}(c) the low temperature thermoelectric current grows with respect to the noninteracting case. For the second parameter set (corresponding to the non-resonant situation) the effect is reversed: the slope of $\widetilde A(\omega)$ decreases and the thermoelectric response of the vibrating molecule is smaller than that of the rigid quantum dot.

For both parameter sets we find that, as $\Delta T\to \infty$, the maximum thermocurrent is largest in the case of zero EP interaction.
This can also be understood from Eq.~(\ref{EQUcurrent}). For large $\Delta T$, the region $\omega\in[-\Delta T,+\Delta T]$ grows and $f_L(\omega)-f_R(\omega)$ is finite far from the Fermi surface. Now the thermoelectric current is given not by the slope of $\widetilde A(\omega)$ at $\omega=0$ but by the ratio of the integrated spectral weight below and above the Fermi surface.
Because $T_P$ is low, the phonon bands in $\widetilde A$ are weighted according to a Poisson distribution with the parameter $\widetilde g^2=\gamma^2\varepsilon_p/\omega_0$. As $\gamma$ grows, a considerable portion of the total spectral weight is shifted to the phonon bands at $\omega > 1$. Even for large temperature differences, their contribution to the thermocurrent will be exponentially small, and the maximum current decreases with respect to the rigid quantum dot.
%
%
\subsection{Thermovoltage}\label{SECcompvol}
\begin{figure}[t]
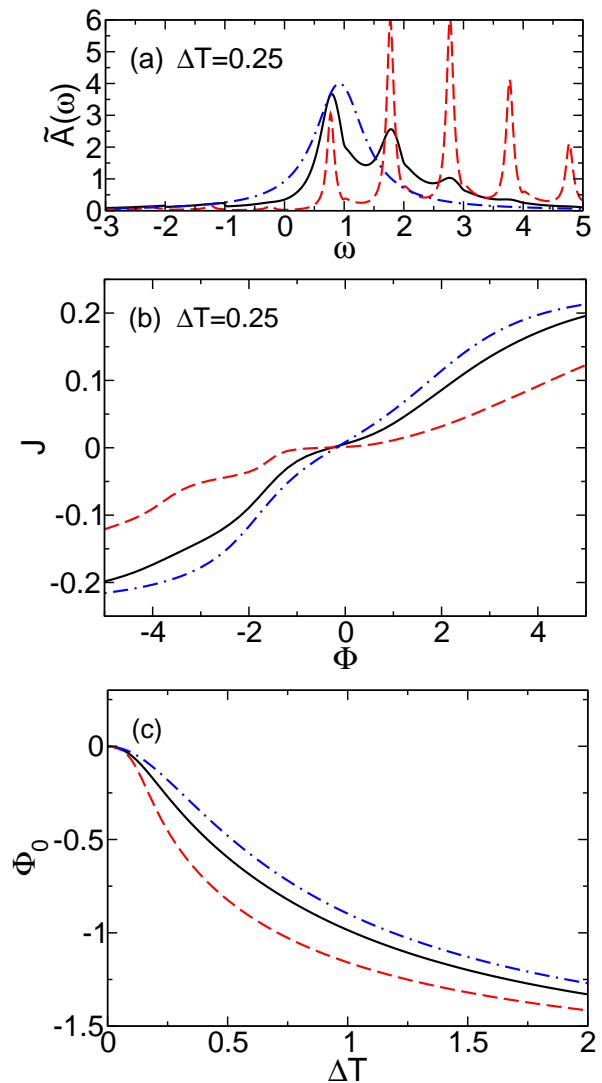

\begin{center}
	\hspace{0.3cm}\includegraphics[width=0.85\linewidth]{Fig3_a}\\[0.2cm]
	\includegraphics[width=0.875\linewidth]{Fig3_b}\\[0.2cm]
	\includegraphics[width=0.9\linewidth]{Fig3_c}
\caption{(Color online) For the same parameters as in Fig.~\ref{FIG2}(b) and (d), i.e., $\omega_0=1$, $T=0.01$, $\varepsilon_p=2$, $\Gamma_0=0.5$, $\Delta=2.5$, $\gamma_{min}=0.54$, $\widetilde\Delta=0.91$ and $\widetilde\Gamma_0=0.28$. (a) Electronic spectral functions for fixed $\Delta T=0.25$ and $\Phi=0$. We compare the variational calculation (black line) to the small polaron limit $\gamma=1$ (red dashed line) and to the result for $\varepsilon_p=0$ (blue dot-dashed line). (b) Current $J$ as a function of the voltage difference $\Phi$ between the leads for fixed $\Delta T=0.25$. (c) Thermovoltage $\Phi_0$ as a function of $\Delta T$.}
\label{FIG3}
\end{center}
\end{figure}
We have seen how, in general, a temperature difference $\Delta T$ between the leads will induce a particle current through the dot. In a typical experiment, this thermoelectric effect is measured by applying a voltage difference $\Phi$ in such a way that the thermally induced current is compensated: $J(\Delta T,\Phi)=0$ for finite $\Delta T$ and $\Phi$. The determination of the so-called thermovoltage, $\Phi_0$, and its dependence on the EP coupling will be the subject of the following numerical calculations. Thereby, we use the second parameter set from Sec.~\ref{SECstate}, i.e. $\omega_0=1$, $T=0.01$, $\varepsilon_p=2$, $\Delta=2.5$ and $\Gamma_0=0.5$, resulting in $\gamma_{min}=0.54$, $\widetilde\Gamma_0=0.28$ and $\widetilde\Delta=0.91$.

To understand the mechanism, we first consider a finite but fixed temperature difference $\Delta T=0.25$. For $\Phi=0$, the corresponding electronic spectral function is shown as the black curve in Fig.~\ref{FIG3}(a), where it is again compared to the noninteracting case and the small polaron limit. We calculate the total current $J$ as a function of the voltage bias $\Phi$, whereby $\mu_L=\Phi/2$ and $\mu_R=-\Phi/2$. The results are depicted in Fig.~\ref{FIG3}(b).

For negative voltages, the current signal is in qualitative agreement with the results of previous works.\cite{Koch2011} In a nutshell, the formation of a polaron like state at the dot reduces the zero-bias conductance due to the Franck-Condon blockade, but for growing voltage we find steps in the current signal whenever $\Phi/2=-\widetilde\Delta-n\omega_0$, with $n\in[0,1,2,\dots]$. Here the chemical potential of the right lead crosses the phonon side bands in $\widetilde A$ and resonant transport of electrons takes place via the emission and subsequent absorption of an equal number of phonons. 
The right lead is the cold one and has a step-like Fermi function. Because of this, the width of the current steps is mainly determined by the width of the phonon bands in the spectral function, i.e., by $\widetilde\Gamma_0$. In the variational calculation the bands overlap considerably (see Fig.~\ref{FIG3}(a)) and the resonant steps in the current-voltage signal are smeared out.

For positive voltage bias we find no current steps even in the small polaron limit. Now it is the chemical potential of the hot left lead that crosses the renormalized dot level and the phonon bands. The width of the resonant tunneling steps in $J(\Phi)$ does not depend on $\widetilde\Gamma_0$ alone but also on the width of the soft Fermi-surface. Since $\Delta T=0.25$ is of the order of $\omega_0$, the current steps are smeared out and therefore no longer discernible. 

Based on the current-voltage signal we now determine the thermovoltage $\Phi_0$ as a function of $\Delta T$ numerically and show the result in Fig.~\ref{FIG3}(c). We find that for small temperature differences, the absolute value of $\Phi_0$ grows strongest for $\gamma=1$. This is a consequence of the reduced electrical conductance of the quantum dot in the small polaron picture; for a given $\Delta T$, relatively large voltages are necessary to compensate the corresponding thermocurrent. 
For large temperature differences, we find a maximum value $\Phi_0=-1.5$ for all three data sets given in Fig.~\ref{FIG3}(c). This voltage corresponds to the postition of the zero-phonon step in $J(\Phi)$ in Fig.~\ref{FIG3}(b). Here, the systems conductance grows considerably and any thermally induced current can easily be compensated by a slight growth in $\Phi$. 

%
%
\subsection{Varying the dot level}\label{SEClevel}
\begin{figure}[ht]
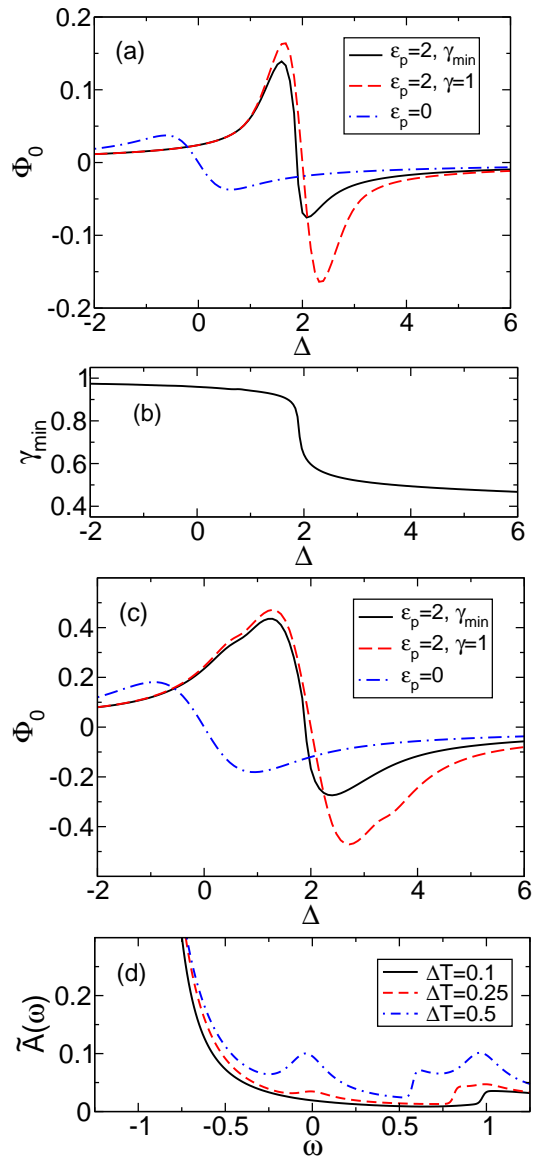

\begin{center}
	\hspace{-0.2cm}\includegraphics[width=0.78\linewidth]{Fig4_a}\\[0.1cm]
    	\includegraphics[width=0.78\linewidth]{Fig4_b}\\[0.1cm]
    	\includegraphics[width=0.8\linewidth]{Fig4_c}\\[0.1cm]
    	\includegraphics[width=0.8\linewidth]{Fig4_d}
\caption{(Color online) For $\omega_0=1$, $T=0.01$, $\varepsilon_p=2$, $\Gamma_0=0.5$, as in Fig.~\ref{FIG3}.  (a) Thermovoltage $\Phi_0$ as a function of the bare dot level $\Delta$ for $\Delta T=0.1$. We compare the variational calculation (black line) to the small polaron limit $\gamma=1$ (red dashed line) and the case with zero EP coupling (blue dash-dot line). (b) Optimal variational parameter $\gamma_{min}$ as a function of the bare dot level. (c) Same as (a) but with $\Delta T=0.25$. (d) Electronic spectral functions for $\Delta=0.8$, $\gamma_{min}=0.95$ and several temperature differences.}
\label{FIG4}
\end{center}
\end{figure}
In the previous sections, we chose the bare dot level $\Delta$ in such a way, that the renormalized level $\widetilde\Delta$ was equal in the scenarios with and without EP interaction. In this way, we concentrated on the influence of the renormalized dot-lead coupling on the thermovoltage. However, in an experimental situation, the dot level may be manipulated, e.g., by applying a gate voltage through a third electrode, in order to optimize the thermoelectric effect. To investigate this situation, we now consider $\Delta$ as our free parameter. For each $\Delta$ we temporarily set $\Phi=0$ and $\Delta T=0$ to determine the optimal parameter $\gamma_{min}$. We then keep $\gamma_{min}$ fixed, set $\Delta T$ to a finite value and calculate the thermovoltage $\Phi_0$. The remaining parameters are the same as in Sec.~\ref{SECcompvol}, i.e.,  $\omega_0=1$, $T=0.01$, $\varepsilon_p=2$, and $\Gamma_0=0.5$.

First, we consider a small temperature difference $\Delta T=0.1$. Our result for the thermovoltage as a function of the bare dot level is presented in Fig.~\ref{FIG4}(a). In general, $\Phi_0(\Delta)$ features two resonances of opposite sign and goes to zero when the renormalized dot level crosses the Fermi surface, i.e., when $\widetilde\Delta=\Delta-\varepsilon_p\gamma(2-\gamma)=0$. At this point the spectral function is symmetrical around $\omega=0$, and the thermoelectric flow and counterflow between the leads cancel exactly. 
As $|\widetilde\Delta|$ grows, the net thermocurrent and, consequently, the thermovoltage increases. In accordance with our results in Sec.~\ref{SECcompvol}, the maximum thermovoltage is largest in the small polaron picture due to the strong renormalization of the dot-lead coupling.
When $|\widetilde\Delta|$ grows further and the zero phonon peak shifts away from the Fermi surface, the dot density of states near $\omega=0$ decreases, and the thermoelectric effect vanishes again.  

In Fig.~\ref{FIG4}(a), the small polaron result is shifted from the $\varepsilon_p=0$ curve by the value of the polaron binding energy $\varepsilon_p$, which can be understood from setting $\gamma=1$ in Eq.~(\ref{EQUdefeta}). In both cases, the thermovoltage signal runs linearly through zero. We note that for growing $T$ (not shown here) its slope reduces, while the position of the positive (negative) resonance shifts to lower (higher) $\Delta$. In this regard, our result resembles the sawtooth-like thermopower signal that was predicted in Ref.~\onlinecite{Beenakker1992} and experimentally measured in Ref.~\onlinecite{Scheibner2008}. There, the periodicity of the thermopower oscillations was determined by the difference in the ground-state energies for different numbers of electrons on the dot. In our model however, we only account for a single dot electron. That is why we only observe a single ``tooth'', i.e., only two thermovoltage resonances in Fig.~\ref{FIG4}(a).

Moreover, we find no side peaks in the thermoelectric signal that could be attributed to the phonon side bands in $\widetilde A$. This is a consequence of the floating effect discussed in Sec.~\ref{SECstate}; since the thermovoltages in Fig.~\ref{FIG4}(a) are small, the expression $f_L(\omega+U_L)-f_R(\omega+U_R)$ in Eq.~(\ref{EQUcurrent}) differs from zero only within the region $\omega\in[-\omega_0+|\Phi_0|/2,+\omega_0-|\Phi_0|/2]$. In this region, the phonon side bands in $\widetilde A(\omega)$ are suppressed and do not contribute to the thermoelectric transport. That is why the resonance signal in Fig.~\ref{FIG4}(a) is given by the varying position of the zero-phonon peak only.

Consequently, in Fig.~\ref{FIG4}(a) the variational calculation with its moderate renormalization of the dot-lead coupling predicts a weaker resonance signal than the small polaron picture, which is also shifted from the result for the rigid dot by less than $\varepsilon_p$. We find that in contrast to the two limiting cases, the strength of the resonance now depends on the sign of $\Phi_0$. This can be understood from Fig.~\ref{FIG4}(b), where we show the optimal variational parameter as a function of the dot level. For $\widetilde\Delta>0$, we have $\gamma_{min}\approx 0.5$ like in the previous sections. When the renormalized dot level crosses the Fermi surface, $\gamma_{min}$ approaches $1$ and a small transient polaron forms at the dot. Then the thermoelectric effect increases.

Figure~\ref{FIG4}(c) shows the thermovoltage signals after having raised the temperature difference to $\Delta T=0.25$. As expected, for all three cases, the heights and widths of the resonances grow with respect to Fig.~\ref{FIG4}(a). More importantly, for finite EP coupling the curves now feature side bands at a distance of $\omega_0$ from their maximum resonances. 
This becomes possible because for large temperature differences the floating condition for the phonon bands is relaxed: Since $\Delta T\lesssim \omega_0$, the Fermi surface of the hot left lead softens considerably. Now the third and fourth lines in Eq.~(\ref{EQUelectronic}) contribute to the spectral function even for $\omega\in[-\omega_0+|\Phi_0|/2,+\omega_0-|\Phi_0|/2]$. For example, Fig.~\ref{FIG4}(d) shows the electronic spectral function in the variational calculation for $\Delta=0.8$ and several temperature differences. Since $\gamma_{min}\simeq 1$, we have $\widetilde\Delta\approx-1.1$. For growing $\Delta T$, a small peak appears near the Fermi-surface that is related to the polaronic state with one phonon at $\omega=\widetilde\Delta+\omega_0\approx0$. When, for varying $\Delta$, this peak crosses the Fermi surface, the side peak in the thermovoltage signal in Fig.~\ref{FIG4}(c) appears. Note, however, that the floating condition still holds for the cold right lead, as can be seen from the shoulder appearing near $\omega=1$ in Fig.~\ref{FIG4}(d). As the thermovoltage grows, the window $\omega\in[-\omega_0+|\Phi_0|/2,+\omega_0-|\Phi_0|/2]$ closes and the shoulder shifts towards $\omega=0$.
%
%
\section{Summary and outlook}\label{SECsummary}
In this work, we investigated the steady-state thermoelectric transport through a vibrating molecular quantum dot in the crossover regime far from the antiadiabatic limit. Within a Kadanoff-Baym formalism that is generalized to account for different lead temperatures, the nonequilibrium dot self-energy was calculated to second order in the dot-lead interaction coefficient. In order to account for the polaronic character of the dot state, we applied a variational small-polaron transformation and determined the degree of transformation by minimizing the relevant thermodynamic potential. 

In essence, we calculated the current induced by a finite temperature difference between the leads. For small temperature differences, the influence of the electron-phonon (EP) interaction strongly depends on the specific system parameters; for a near-resonant dot level, the interacting quantum dot acts as a more efficient energy filter and the thermocurrent increases. In the tunneling regime, however, we found a reduction of the thermocurrent due to the decreasing density of states near the Fermi surface.

In order to relate our results to experiment, we determined the thermovoltage required to compensate the thermally induced particle current at a given temperature difference. We found that in principle the Franck-Condon blockade boosts the thermovoltage through the reduction of the systems electrical conductance. For intermediate EP coupling, the small polaron picture overestimates this effect.

Finally, we determined the thermovoltage as a function of the dots energy level. Because of Pauli blocking, we found no phonon features in the low-temperature thermoelectric signal. However, our variational calculation predicts an asymmetrical line shape due to the formation of a small polaron as the dot level drops beneath the equilibrium Fermi energy. For large temperature differences of the order of the phonon energy, Pauli blocking is relaxed and the thermovoltage signal features multiple resonances that can be attributed to resonant transport through vibrational dot states.

The present study should be considered a first step in applying our variational ansatz to the thermoelectric transport through molecular junctions. Although it captures the essential physics, it must be extended in several directions. 
Most importantly, we have yet to consider energy transport through the junction. The efficiency of energy deposition by the excitation of local phonons will strongly depend on the effective EP interaction, i.e., on the polaronic character of the dot state. To investigate the subsequent heating or cooling of the molecule, the effective temperature of the dot electrons has to be determined numerically, for which several methods have been proposed.\cite{Galperin2007b} This should have some influence on the variational parameter $\gamma$ as well. In our previous works, a voltage-dependent variational parameter was responsible for the junctions negative differential conductance. Therefore, it would be highly desirable to carry this nonlinear behavior over to the discussion of thermoelectric transport.

Another worthwhile extension of our work concerns Coulomb interaction effects.  Recently, Andergassen {\it et al.}\cite{Andergassen2011} argued for the negative--$U$ Anderson model (which neglects the coupling to the phonon degrees of freedom however) that the resulting charge Kondo effect leads to a large enhancement of the linear response thermopower due to the highly asymmetric dot spectral function.  This effect was shown to be tunable applying a gate voltage. It would be interesting to reexamine this problem for a model with additional EP  interaction, using our variational scheme, since---as we have demonstrated in Sec.~\ref{SEClevel}---the effective EP interaction (described by the variational parameter) is strongly  influenced by the gate voltage. In this connection we like to stress that for a combined Holstein-Hubbard quantum dot model it has been shown that strong EP coupling may result in a net attractive Coulomb interaction.\cite{Cornaglia2004} Then, depending on the energy of the dot level, a variational ansatz might be able to interpolate between the positive--$U$ Holstein-Hubbard dot model and the effective anisotropic Kondo model regime investigated in Ref.~\onlinecite{Andergassen2011}.

\begin{acknowledgments}

This work was supported by Deutsche Forschungsgemeinschaft through SFB 652 B5. TK and HF acknowledge the hospitality at the Institute of Physics ASCR.

\end{acknowledgments}


%
\end{document}